# A Linear Programming Driven Genetic Algorithm for Meta-Scheduling on Utility Grids

Saurabh Garg, Pramod Konugurthi and Rajkumar Buyya

*Grid Computing and Distributed Systems Laboratory, CSSE*
*Department of Computer Science and Software Engineering*
*The University of Melbourne, Australia*
{sgarg, pramod, raj}@csse.unimelb.edu.au

*Abstract*—The user-level brokers in grids consider individual application QoS requirements and minimize their cost without considering demands from other users. This results in contention for resources and sub-optimal schedules. Meta-scheduling in grids aims to address this scheduling problem, which is NP hard due to its combinatorial nature. Thus, many heuristic-based solutions using Genetic Algorithm (GA) have been proposed, apart from traditional algorithms such as Greedy and FCFS.

We propose a Linear Programming/Integer Programming model (LP/IP) for scheduling these applications to multiple resources. We also propose a novel algorithm LPGA (Linear programming driven Genetic Algorithm) which combines the capabilities of LP and GA. The aim of this algorithm is to obtain the best meta-schedule for utility grids which minimize combined cost of all users in a coordinated manner. Simulation results show that our proposed integrated algorithm offers the best schedule having the minimum processing cost with negligible time overhead.

## I. INTRODUCTION

Grid computing enables harnessing of a wide range of heterogeneous, distributed resources for solving compute- and data-intensive applications [23]. Recently, it has been rapidly moving towards a pay-as-you-go model wherein providers expect an economic compensation for the computational resources or services offered to grid users. In one side there are users with applications to execute and, in the other side, there are providers [7] willing to offer their resources or computing services in return for regular payments. Environments with this decoupling of users from providers are generally termed as utility grids [24].

Scheduling in utility grids is complex due to the distributed ownership of the resources. Moreover, consumers and providers are independent from one another and have different access policies, scheduling strategies and objectives [16]. Previous work has proposed grid market infrastructures [25][14] [15] for utility grids. Although this work provides the basis for resource markets, application scheduling considering aspects such as the distributed resource ownership and cost minimization under these scenarios is still in its infancy. Furthermore, existing Grid brokers, that consider cost minimization, are generally single-user based [17][18][25]. These single-user brokers may lead to sub-optimal schedules and are certainly not designed with the aim of minimizing the cost for a group or community of grid users.

In this work we focus on meta-scheduling of different applications from a community of users considering a commodity market. As an example of a user community, we consider the financial institution Morgan Stanley that has various branches across the world. Each branch has computational needs and QoS constraints that can be satisfied by grid resources. In this scenario, it is more appealing for the company to schedule various applications in a coordinated manner. Furthermore, another goal is to minimize the cost of using resources to all users across the community (the company in this case). This meta-scheduling problem is NP hard due to its combinatorial nature.

This paper proposes a novel LP driven meta-scheduling algorithm that maps independent applications with a given deadline and budget to rented resources in the grid with the aim of minimizing the combined cost of all the users who share a meta-scheduler. We have combined the benefit of LP and GA to obtain an efficient solution for the scheduling problem in the grid with concurrent users and multiple heterogeneous resources. Although only three QoS parameters are considered (i.e. number of processors, deadline and budget), the proposed algorithm is general enough to deal with more parameters. A simulation study shows the effectiveness of the proposed algorithms compared to other greedy and genetic algorithms.

The main contributions of this paper are: (a) an LP/Integer Programming (IP) model for the meta-scheduling problem; (b) Modified Mincost (MMC) algorithm for approximating an LP/IP optimal solution for those applications that can run on only one grid resource (single-grid node) and (c) LPGA for scheduling single-grid node applications.

The rest of the paper is organized as follows. In the next section, we describe the related work on meta-schedulers and heuristic based scheduling. In Section 3, we present the problem definition, which is given with the LP model of the scheduling problem. Then, in Section 4, we present LPGA scheduling algorithm. In Section 5, we discuss simulation study of various algorithms and their comparison with the proposed algorithms. Finally, Section 6 concludes the paper and described the future work.

## II. RELATED WORK

Market based mechanisms can be divided into two based on market models i.e. auctions and commodity market model. As our work is relevant for commodity markets in Grid thus in this section we will compare our with other resource allocation mechanisms for this market model. Cost-based resource management and market-oriented models have proven to be useful for resource management [18]. The most relevant works to our research are [9][19][21] and [20]. In our previous work Gridbus Broker[9] and Nimrod/G[25], a greedy approach is proposed to schedule a parameter sweep application with deadline and cost constraints. To be precise, our previous work on grid scheduling [9][25] was focused on application level scheduling, i.e. a personal broker for efficient deployment of an individual application on utility grids. But this work is focused on scheduling of many applications from users having different QoS requirements with the aim of global optimization across many user applications.

G-commerce [19] is another economic-based study that applies strategies for pricing grid resources to facilitate resource trading, and it compares auction and commodity market models using these pricing models. Feng [21] proposed a deadline cost optimization model for scheduling one application with dependent tasks. These studies has many limitation (1) algorithms proposed are not designed to accommodate concurrent users competing for resources (2) also application model considered are simple i.e. independent task or parametric sweep application. In this work we have modeled both independent and parallel applications submitted by concurrent users. Similarly, Dogan [20] proposed a metascheduling algorithm considering many concurrent users but application model was again very simple. They assumed that each application consists of one task and also each application is independent. In this paper, we have considered multiple and concurrent users competing for resources in a meta-scheduling environment to minimize combined cost of all users. Moreover, in this paper, we presented a formal theoretical mathematical model for optimal scheduling of jobs with Deadline and Budget constraints which is essential to develop an effective heuristic that may provide a near optimal solution.

In the grid environment, where each user has different QoS constraints, the scheduling problem becomes a Generalized Assignment Problem (GAP) which is a well known NP hard problem. GAP can be solved using heuristic based GA [2][4]. Thus many GA based heuristics are proposed in the literature Wiessman et al. [6] proposed a novel GA based algorithm which schedules a divisible data intensive application. Martino et al. [10] presented a GA based scheduling algorithm where the goal of super-scheduling was to minimize the release time of jobs. These GA based heuristic based solutions [13][10][6] do not consider QoS constraints of concurrent users such as budget and deadline. The viability of evolutionary algorithm-based scheduling such as GA for realistic scientific workloads is demonstrated by systems like Mars [11]. Mars is a meta-scheduling framework for scheduling tasks across multiple resources in a grid. In a recent study [5], five heuristic based algorithms were compared, which found that GA produces the best utility for users in comparison to heuristics such as Min-Min. Due to these reasons we have used GA seeded with greedy as an algorithm to compare with.

From scheduling approaches outside the Grid computing, it is well known that LP/IP can offer optimal solutions in minimization/ maximization scheduling problems when constraints and objective functions are linear. In Feltl et al. [4], an LP/IP based initial solution and followed by an intelligent replacement of offspring based on Martello et al. [3] is proposed. But these models do not consider the deadline and budgets constraints. As they are developed based on specific domain knowledge, they can't be applied directly to grid scheduling problems, and hence have to be enhanced accordingly.

The main contribution of this paper is to develop a mathematical model and an LP/IP based GA algorithm to minimize the combined procession cost of user application in a concurrent user's environment for utility Grids. To best of our knowledge, the solution methodology proposed doesn't overlap with existing techniques for scheduling problems.

## III. PROBLEM DEFINITION

### A. System Model

Figure 1 shows the interaction of the MetaBroker with resource providers and users. Grid users submit their application jobs to the MetaBroker with their

QoS requirements. The QoS requirements consist of budget, deadline, and number of processing elements (PEs) i.e. CPU required. Each application job consists of independent tasks with each task requiring one PE. A resource comprises of homogeneous PEs. Two kinds of user's application jobs considered are:

- Multi-grid node (MGN) jobs can split into independent tasks, which can run on different resources. For example, bag of tasks applications.
- Single-grid node (SGN) jobs which require a complete set of PEs on the same resource and fail if all PEs are not available. For example, synchronous parallel applications.

The MetaBroker gathers information about resources such as PEs available, grid middleware, cost of PEs per unit time for each user and PEs capacity (e.g. Millions of Instruction per Seconds (MIPS) rating) from the resource providers who have agreed to rent their computational resources to MetaBroker. The MetaBroker generates a combined schedule based on algorithms discussed in subsequent sections. As the demand for resources may exceed the supply in the grid, some jobs are placed in the queue for latter assignment when resources are available or freed.

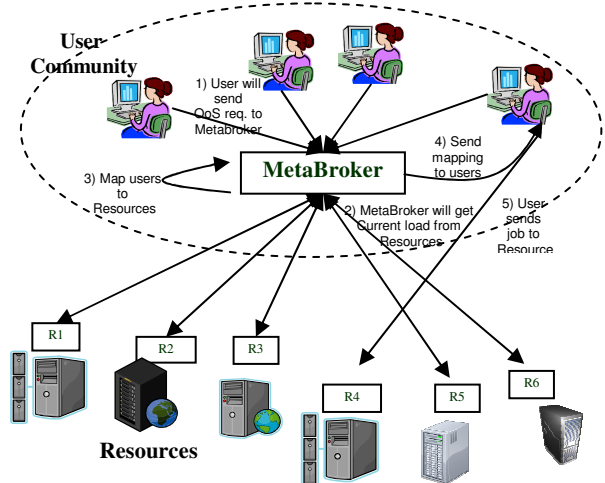

**Figure 1. Grid model with MetaBroker**

### B. Problem Formulation

The objectives of meta-scheduling in the above environment with $m$ resources and $n$ jobs are as follows:

- Scheduling based on QoS, i.e., budget and deadline.
- Minimize combined cost for all the users by meeting the budget and deadline constraints.
- Maximize the total number of user's job submission.

Let $P_i$ represent the information which MetaBroker receives from grid resource $i$ at a given instance $T$.

$P_i : (r_i, n_i, c_{i,j}, v_i)$ where

$r_i$ is the Resource Id,
$n_i$ is number of free PEs available on the resource,
$c_{i,j}$ is the cost of using a single resource per second for the job/user $j$,
$v_i$ is the MIPS speed of one of the PE,
and $i \in I = \{1..m\}$

Let $Q_j$ represent the QoS demand which MetaBroker receives from the user $j$ at a given time instance $T$.

$Q_j: (u_j, j_j, b_j, d_j, M_k, m_j)$ where

$u_j$ is the user id,
$j_j$ is the job id, since one user can submit multiple jobs, hence combination $(u_j,j_j)$ is unique,
$b_j$ is the budget constraint the user specifies,
$d_j$ is the deadline constraint for finishing the job,
$M_k$ size of each task in the job in terms of Millions of Instructions (MI) [22][9],
$m_j$ is the number of PEs the user requires for executing the job,
and $j \in J = \{1..n\}, k \in K = \{1..m_j\}$.

The mathematical model for scheduling of SGN jobs is as follows:

$$c = \sum_{i=1}^{m} \sum_{j=1}^{n} c_{ij} x_{ij} r_{ij} Max_k (M_k/v_i) \quad (1)$$

In Equation (1), the variable $r_{ij}$ denotes the number of PEs allotted to a job $j$ on a resource $i$. Note that, in case of SGN jobs, the value of $r_{ij}$ would be either 0 or $m_j$ (the number of PEs required by the job $j$). The variable $x_{ij}$ denotes that whether job $j$ is mapped to resource $i$ or not. Equation (1) denotes cost function which all users have to pay to resource providers. The constraints are following :

$$\sum_{j=1}^{n} r_{ij} x_{ij} \leq n_i, \forall i \in I \quad (2)$$

$$\sum_{i=1}^{m} r_{ij} x_{ij} = m_j, \forall j \in J \quad (3)$$

$$x_{ij} \in \{0,1\}, \forall i \in I, j \in J \quad (4)$$

$$\sum_{i=1}^{m} x_{ij} = 1, \forall j \in J \quad (5)$$

$$\sum_{i=1}^{m} r_{ij} c_{ij} \leq b_j, \forall j \in J \quad (6)$$

$$MAX(M_k / v_i) x_{ij} \leq d_j, \forall i \in I, j \in J, k \in K \quad (7)$$

$$c_{ij}, r_{ij}, b_j \text{ and } d_j \geq 0, i \in I, j \in J \quad (8)$$

Equation (2) denotes the resource's capacity constraints. Equation (3) denotes the PE requirement of user's job i.e., $m_j$. Equations (4) and (5) denote the assignment matrix, which indicates that a job can be assigned to a maximum of one resource. Equation (6) is added for satisfying the budget constraints and Equation (7) is added for the time constraints. Equation (1) is solved for minimization to obtain an optimal cost efficient schedule. All the quantities, i.e., the resources, cost, deadline and budget are greater than zero, hence the Equation (8). Note that Equations (4) and (5) make the problem NP-hard as these constraints make the problem a 0-1 assignment problem which is well known to be NP-hard [3][4][8].

LP/IP model for scheduling MGN jobs is almost same as for SGN jobs. The only difference is that model for MGN jobs will not include constraints (4) and (5).

### C. Genetic Algorithm Formulation

Genetic Algorithms (GAs) [8] provide robust search techniques that allow a high-quality solution to be derived from a large search space in polynomial time, by applying the principle of evolution. A genetic algorithm combines the exploitation of best solutions from past searches with the exploration of new regions of the solution space. An individual (chromosomes) represents any solution in the search space of the problem. A genetic algorithm maintains a population of individuals that evolves over generations. The quality of an individual in the population is determined by a fitness-function. The fitness value indicates how good the individual is compared to others in the population. A typical genetic algorithm consists of the following steps:
 (1) Initial population generation;
 (2) Evaluation;
 (3) repeat
 (4) Selection;
 (5) Crossover;
 (6) Mutation;
 (7) Evaluation;
 (8) until (convergence).

Convergence can be a different criterion for different problems, but generally 'a no change in the solution for *n* generations' is considered as convergence. The most important aspect in GA is the solution space encoding and the fitness function. We have taken the same fitness function and solution space encoding as given in our previous work [8]. The genetic operations, i.e., crossover, mutation and inversion are standard operations in GA.

### IV. PROPOSED ALGORITHMS

#### A. Linear Programming based Algorithm for Scheduling MGN Jobs

As discussed in the previous section, the mathematical model for MGN jobs scheduling will be converted to an LP/IP model after removing constraints (4) and (5). LP/IP can be then easily solved using standard integer programming algorithms. For simulation purposes, we have used the Coin-OR library [12].

#### B. Linear Programming based Algorithm for Scheduling SGN Jobs (LPGA)

Due to the absence of constraints (4) and (5), LP/IP solutions obtained in section 4.1 are infeasible for the SGN jobs scheduling problem. We need to approximate LP/IP to a feasible solution. For this problem, we have designed a novel algorithm MMC to approximate an infeasible LP/IP solution to a feasible solution for the SGN jobs scheduling problem.

*1) Modified MinCost (MMC) Algorithm:* Our MMC algorithm is inspired from Minimum Cost Algorithm which is a well-known method for finding the feasible solution near to optimal in the transport problems. In the Minimum Cost algorithm, we find the minimum cost resource, assign maximum number of jobs to that and reduce the demand and supply. The process is repeated until all jobs are allocated.

In MMC, we take an optimal solution from the LP based algorithm (discussed in the previous section) and then approximate it to obtain a feasible solution for SGN scheduling problem. This algorithm gives a solution, which is between the optimal, and the greedy solutions, i.e. it will give a better solution than greedy but not an optimal as shown in the evaluation section. The solution obtained from this algorithm acts as an initial solution to GA.

The pseudo code for MMC is given in Algorithm 1, which shows how MMC approximates an LP/IP solution. It takes as input a list of jobs which is a structure containing information regarding how different tasks of a job are mapped to resource providers and how many PEs are required by the job. This job list is generated from algorithm for scheduling MGN jobs. First, MMC algorithm makes a copy of the input, and then sorts job list on the basis of number of resources to which the job is mapped (line 1-2). This is done so that we get a solution as closer as possible to the optimal LP/IP solution.

We handle differently the jobs which are mapped only to (a) one resource and (b) more than one resource. For jobs which are mapped only to one resource, we freeze their mappings. The mappings will not change during the algorithm, and thus we reduce the total capacity of resource (line 4, 5, 6). Then, these jobs are added later to the schedule list (line 7). The jobs, which were mapped to more than one resource, will be mapped to a resource $p_i$ which has the capacity to allocate the full job $j1$ and is most economical (line 9-11). Other jobs which are allocated to this resource will be remapped to resource where job $j1$ was allocated other than $p_i$, if deadline and budget constraints are satisfied; otherwise they are allocated to dummy resources (line 15). Other jobs, which are not allocated in the above process, will be mapped to the dummy resource (line 19). These jobs are remapped to real resources using a greedy approach.

**Algorithm 1:** ModifiedMinCost *(Listjobs)*

```
1   ListJ←Listjobs.clone();
2   sort(ListJ);
3   for i←0 to ListJ.size()-1 do
4       j1←ListJ(i);
5       if (numofProviders(j1)==1) then
6           changeAvailableCapacity(Provider(j1),j1);
7           Goto step 20
8       else
9           P←J1.providerlist.clone();
10          sort(P); //number of PE allocated on
                     Provider
11          for p←0 to P.size() do
12              if(remainingCapacity(p)>PERequired(j1))
13                  AllocateFull(j1,p);
15                  InterchangeCapacity(p,p.joblist[]);
16                  Goto step 20
17              endfor
18          endif
19          addjob(j1,DummyResProv);
20      endfor
21  ScheduleDummyResJob();
22  MakeScheduleList(ListJ);
23  return ListJ;
```

*Complexity:* Complexity of MMC algorithm for m resources and n jobs is $O(nlogn + mn + O(interchangeCapacity() function)*n)$. We have taken a simple implementation of interchangeCapacity function using arrays, which has worst-case complexity as $O(mn)$. Thus, the worst-case complexity of the algorithm becomes $O(mn^2)$, which is very high upper bound over real complexity as average number of jobs per resource will be less than total number of jobs.

*2) LPGA Algorith:* LPGA seeds the approximate LP solution with Genetic algorithm for generating a solution near to optimal solution. In LPGA, first we solve an LP/IP problem without constraints (4) and (5) and then MMC approximates the solution obtained in the previous step. Finally, using the feasible solution from MMC, we iterate various genetic operations to get the best solution.

**Algorithm 2.** LPGA()

```
1   ScheduleList←null;
2   while (current_time < next_schedule_time) do
3       RecvResourcePublish(Pj); //from providers
4       RecvJobQos(Qj); //users
5   endwhile
6   if (numJobs>numResource) AddDummyNode();
7   ListP←ListProviders.clone();
8   Sort(ListP) //cost of resource
9   for each job 'j' in Q do
10      compute QoS index
11  endfor
12  ListJ=sort(Q); //descending order of QoS index
13  LPschedule=GetLPsolution();//from Coin Library
14  ScheduleList= ModifiedMinCost(LPschedule);
15  AddinPopulation(POPU_LIST, ScheduleList);
16  generateNextGenerationPopulation(){
17      for each i in (population_size-POPU_LIST.size())
18          cromos=GenerateRandomCromosome(i);
19          POPU_LIST.add(cromos);
20      endfor
21      ComputeFitnessfunction();
22      bestCromos=SearchBestCromosome(POPU_LIST);
23      if (termination) return bestCromos
24      doSelection(){
25          SelectBestCrossoverRateCromosome() // 'based
                on 'Roulette wheel selection policy'
26          DoCrossover()
27          DoMutation()
28          AddtoPopulationList(POPU_LIST);
29          Goto step 16
30          }
31  }
32  Schd_List =GenerateSchedulelist(bestCromos);
33  for each element in Schd_List do
34      notifyuser();
35  endfor
```

Algorithm 2 gives the pseudo code. As the MetaBroker at a particular time instance does scheduling, it waits for jobs from users (taken as QoS requirement for job in line 4) and resource load from provider (line 3). Then we have to add dummy job queue with high cost and number of PEs to make scheduling problem balanced, i.e. to balance supply of resource service and demand by users for computing services (line 6). After that, we sort resource list on the basis of their cost and job list on the basis of QoS index (line 7-12). After sorting resource list and job list, we get the LP/IP solution which is obtained without considering constraints (3) and (4) discussed in section 3.2 (line 13). This solution is approximated using MMC algorithm (line 14) which is used to generate initial population space for the genetic algorithm (line 17-20). From initial population space, we select the best chromosome values based on the fitness function. These best chromosomes are then mutated to generate the next generation population (line 21-30). This operation is repeated until the convergence criterion is reached or maximum specified limit of iteration is reached (line 23). Based on the chromosome, schedule list is generated and users are notified about mappings (line 32-34). One limitation to the problem formulation is that if the number of resources requested is more than the number of resources available for the broker, then the problem results in an infeasible solution. Therefore, we propose addition of dummy resources with infinite capacity and having more cost than any of the resources available on the grid. This enables the algorithm to converge and assign some of the jobs to the dummy resources. Dummy resource jobs are rolled over to the next schedule period, as they cannot be executed.

## V. PERFORMANCE EVALUATION

### A. Simulation Methodology

We use the GridSim toolkit [1] to simulate a grid environment and the MetaBroker for scheduling user's application. We simulated the grid with heterogeneous resources having different MIPS [22] ratings and each resource having different number of PEs in the range of 4 to 12 with mean number of PEs as 8. A Grid having resources in the range of 25 to 200 is generated. The usage cost per second of each resource is varied (using Gaussian distribution) between 4G$-5G$ (G$ means Grid dollar) with mean cost of 4.5G$. Execution time in the simulator setup for the several hundred runs restricts us to model 50 jobs per run. Thus, the MetaBroker schedules the jobs (which consist of several tasks, each require one PE to run) submitted by 50 concurrent users with simulation interval of 50 seconds. The jobs are also simulated with varying QoS requirements. Jobs with average number of 5 tasks and having 10-50% variations in the number of tasks (using Gaussian distribution) are considered and all the jobs are submitted within 20 seconds of simulation start time.

Average estimated run time for jobs is also taken to be 400 seconds and varied by ±20% using Gaussian distribution. As far as QoS parameters are concerned, i.e.., budget and deadline, three simulations are done for different deadline scenarios as listed below:

**Experiment-1**: Tight deadline (estimated time+(50 seconds with ±20% variation)).
**Experiment-2:** Medium deadline (estimated time+(250 seconds with ±20% variation)).
**Experiment-3:** Relaxed deadline (estimated time+(500 seconds with ±20% variation)).

As far as the budget is concerned, all the jobs are given relaxed budget constraints (i.e., twice the cost of average job execution time). Our performance evaluation examines the relative performance (users spending) of LPGA with respect to 3 other following meta-scheduling algorithms for the above type of jobs by varying the number of resources:
1. Greedy based meta-scheduling algorithm.
2. Heuristic GA algorithm (HGA) [8].
3. Modified Mincost algorithm (MMC).

The results for the three different simulation scenarios for both MGN LP based algorithm and SGN LP driven Genetic Algorithm (LPGA) are discussed in the next section.

### B. Performance Results

*1) LP/IP based Meta-Scheduling Algorithm for MGN Jobs:* The results are compiled in Table 1. In tight deadline out of 50 jobs (246 tasks) only 34 jobs (170 tasks) were scheduled where as in medium and relaxed deadlines all jobs i.e. 50 (246 tasks) were scheduled. The reason behind this is the tight deadline, for which many jobs missed the deadline constraints due to simulation interval. Thus, even though the number of resources increased in number, the number of scheduled jobs remained same.

TABLE 1.
TOTAL COST SPENT BY USERS FOR MGN JOBS

| Number Of Providers | Total Cost in Medium Deadline | Total Cost in Relaxed Deadline | Total Cost in Tight Deadline |
|---|---|---|---|
| 25 | 259832.6 | 194668.9 | 245280.8 |
| 50 | 261059.7 | 198771.5 | 247655 |
| 100 | 260689.1 | 198462.5 | 247251.2 |
| 150 | 260653.5 | 198360.5 | 247245.4 |
| 200 | 260664 | 198455.3 | 247015.7 |

Moreover, in case of relaxed deadline, user spending is the minimum in comparison to other deadline types because more jobs are scheduled to resources with least cost. Also, even though it seems that this trend is not followed when we compare tight deadline and medium deadline scenarios, it results in more number of jobs scheduled in the case of medium deadline. It is also interesting to note that in the tight deadline scenario, with the increase in number of resources; total cost spent by users is also decreasing.

*2) SGN Job Scheduling Algorithms:* The performance of four meta-scheduling algorithms has been compared by varying the number of jobs scheduled and users spending. The results are compiled and presented in Fig. 2 to Fig.4. In the first set of results, i.e., Fig. 2a, 3a and 4a, aggregated revenue earned for the completed jobs are plotted on Y-axis. In the first set of results, i.e. Fig. 2b and 3b, X-axis presents the grid resources while Y-axis presents the jobs completed (out of the 50 concurrent jobs submitted). Due to space constraint we have not presented the number of jobs completed with number of resources and also trend in medium deadline case is very similar to trend in case of relaxed deadline.

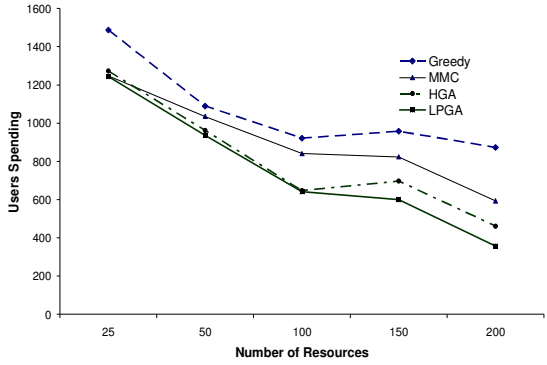

(a)

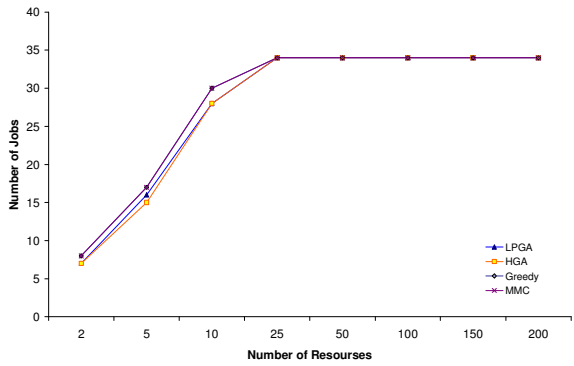

(b)

**Figure 2. Tight Deadline**

The figures 2a, 3a and 4a for relaxed, medium and tight deadlines show that LPGA has outperformed all other meta-scheduling algorithms, thus minimizing the users spending. In case of all deadline, even though initially, user spending in schedule given by HGA and LPGA seems to be same, as number of resources increases, LPGA decreases the user spending if compared to other algorithms. Since the cost variations within the grid resources are not significant (i.e. 4.5 G$ with ±0.5G$) only up to 7-8 % cost benefit was noticed. However, more benefit can be anticipated if the variations are higher.

More interesting results can be observed when we compare performance of MMC algorithm and greedy approach. It is clear from the graphs that cost gain for users in case of MMC algorithm can be as much as 30% more beneficial for users than greedy approach. Thus, the results demonstrate the user spending for each job decreases when LPGA is used by Meta-Broker for combined scheduling of jobs. This is because MMC algorithm is giving better schedule than greedy algorithm.

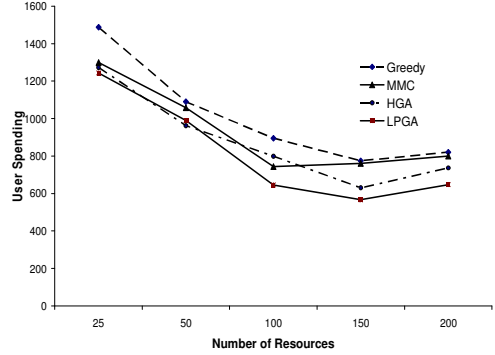

(a)

**Figure 3. Medium Deadline**

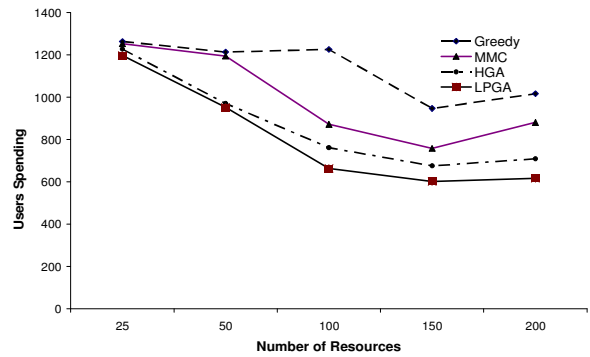

(a)

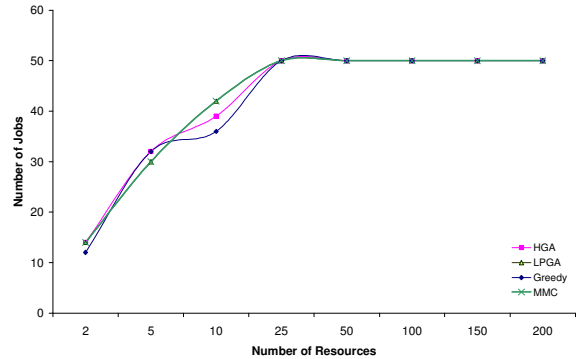

(b)

**Figure 4. Relaxed Deadline**

It can be noticed from Fig. 2b and 4b that the curves become smoother and peak early compared to tighter deadline. This trend is due to the relaxed deadline having more jobs completed with the same number of resources. As when deadline is more relaxed, meta-brokers try to complete more and more tasks. With increase in number of resources, all algorithms schedule the same number of user jobs. In case of tight deadline (fig. 2b), as many jobs missed the deadlines (very tight); only 34 jobs are completed even though the number of resource increased to 200.

Next simulation study presents the advantage of LPGA algorithm over HGA algorithm. We have compared the number of iteration taken by both algorithms in case of tight deadline. The results shown in Fig. 5 indicate that in case of LPGA, number of iterations is reduced with a decrease in user spending. For example, with 50 resources, the number of iterations performed in HGA is 1000, whereas in LPGA it is less than 700. Even though the number of iterations in HGA is less than in LPGA with 150 resources, the revenue generated is smaller which can be observed from Fig. 3b. These results also indicate that in spite of taking more iteration, the HGA could not find the global/better optimum. Hence, to find a better solution (either the one found by HGA implementation or any other better solution) the algorithm would have required a higher number of iterations. This clearly explains the reason for a higher number of iterations in case of LPGA when the number of resources is 150.

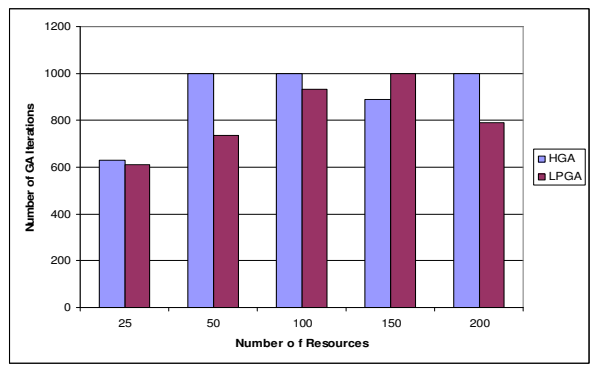

**Figure 5. Comparison of number of iterations in HGA and LPGA**

## VI. CONCLUSION AND FUTURE WORK

In this paper, we have modeled the scheduling concurrent user jobs in utility grids as an LP/IP problem. Two different types of jobs were considered, namely multi-grid node jobs and single-grid node jobs. We also presented a novel algorithm called MMC that approximates the LP solution to get a single-grid node job scheduling solution near to optimal. We designed the LPGA algorithm to decrease the combined user spending and resource utilization by seeding the solution from MMC algorithm to genetic algorithm. We have also taken into consideration user QoS constraints, i.e., deadline, number of PEs and budget.

The results show that LPGA out performs other meta-scheduling algorithms such as greedy approach and HGA by giving the minimum processing cost. In LPGA, not only number of iterations is reduced by 7% to 25% compared to HGA, but user spending is also decreased as the number of resources increases. Our results also show that MMC can reduce the combined user spending up to 30% over traditional greedy approach.

The application of approximation algorithms like MMC in Grid environment is novel. In addition, this work shows us that many good heuristics can be designed after some modifications. In future, our goal is to study and analyze such algorithms. Also we would like to integrate this algorithm with some available meta-schedulers and deploy them on real test beds. We have given priority to jobs that are mapped to minimum number of resource providers in the LP solution. For future work, we intend to improve the performance of MMC and test it with different priority indexes such as deadline and execution time. We also want to investigate other approaches used for the assignment problem which can be modified for utility grids. We want to experiment our algorithm with a wide range of costs to study the effect of costs change.

ACKNOWLEDGMENT

We would like to thank our colleagues – Marcos, Marco Netto, Srikumar Venugopal and Chee Shin Yeo – for their comments and suggestions on this paper. This research is funded by the International Science Linkage grant provided by the Department of Innovation, Industry, Science and Research (DIISR) and Australia Research Council (ARC) of the Australian Government.